# Challenging the Eye-Mind Link Hypothesis: Visualizing Gazes For Each Programming Problem

**Michael T. LOPEZ II**
*Ateneo de Manila University, Philippines*
michael.lopez@student.ateneo.edu

**Abstract:** This investigates the relationship between eye fixation patterns and performance in Java programming exercises using eye-tracking technology. Thirty-one students from a university in Metro Manila participated, and their eye movements were recorded while solving five Java programming exercises (three of the five exercises were picked). The fixation data were preprocessed and visualized using heatmap bin graphs, dividing the participants into correct and wrong answer groups. The Mann-Whitney U Test was employed to determine if there were significant differences in the fixation patterns between the two groups.

**Keywords:** eye-tracking data, eye fixation, heatmap visualization, problem-solving, programming

## 1. Introduction and Related Literature

Just and Carpenter (1980) developed the foundation of interpreting eye-tracking data. They posited there is "no appreciable lag" between the object being observed and the person processing it. This "immediacy" gives the person's power to instantaneously interpret the subject (Sharafi et al., 2020). It goes well with the eye-mind hypothesis, an idea suggesting that the fixations are synonymous to understanding (Mayer et al., 2023). However, these two assumptions have been recipients of uncertainty. Practices such as passively staring to waste time is an example that is counterintuitive in respect to learning (Reichle et al., 2010).

On gauging the mental effort while reading mathematics textbooks, Molina and colleagues (2018) used two eye-tracking metrics: total fixation counts, which refer to the overall number of fixations generated across all Areas of Interest (AOIs), and fixation density, which represents the concentration of fixations on specific AOIs. Their study discovered that adding more visual designs on the learning materials decreases the cognitive load for processing information. On the other hand, studies have been frequently comparing between low-performing and high-performing groups. Tablatin and Rodrigo (2023) suggested that high-performing students practices more effort on reading the compiler error messages. Similarly, the high performing groups are also efficient in terms of fixing the programming code. This implies that the high performers have more of an "analytical mind" (Pacol et al., 2023).

## 2. Methodology

### 2.1 Sources of Data

This is a spin-off from Pacol et al. (2023), which utilized the same dataset and Java programming exercises. As an overview, we focused on the 31 student-participants from School A in Metro Manila. Each student solved the problem individually. Participant No. 12 was removed due to the lack of recorded fixations during their eye-tracking session. Each participant has a comma separated value (CSV) file of their respective fixation locations, timestamps, durations, blinking counts, and pupil dilations.

## 2.2 Data Preprocessing and Conversion of Data into Graphs

The eye-tracking data was preprocessed by converting the fixation durations from seconds to milliseconds. Rectangles were encoded in the background to visualize on what specific part of the code was given importance. The rule of what each color means are the following: red represents the error line in the code, dark green represents the error compiler message, yellow represents the Java imports, pink represents the beginning and end of the class definition, and the light green represents the beginning and end of the main method.

All 30 individual CSV files were divided into two groups: those who answered it right and wrong for that respective programming exercise. Exercise 1 (18 got it correct), a calculator program that computes addition, subtraction, multiplication, and division; Exercise 3 (17 got it correct), a program that determines whether the string is a palindrome; and Exercise 5 (20 got it right), a rock-paper-scissors command-line game, were the programming codes considered. Only one person answered Exercise 2 wrong, while Exercise 4 was incorrectly solved by four participants. Thus, both Exercise 2 and 4 were excluded. The graphs were separated into 20 by 20 bins. Its purpose is to visually show on how concentrated the fixations are for each part of the screen and per problem. Each of the bins are represented by a shade of blue. This means, the darker the blue color is, the more fixation data that resides inside the bin. On the right side of the heatmap, there is a y-axis labelled for rows 0 to 19.

## 3. Results of the Fixation Point Graphs

### 3.1 Visual Cumulative Fixation Point Graph of Exercise 1

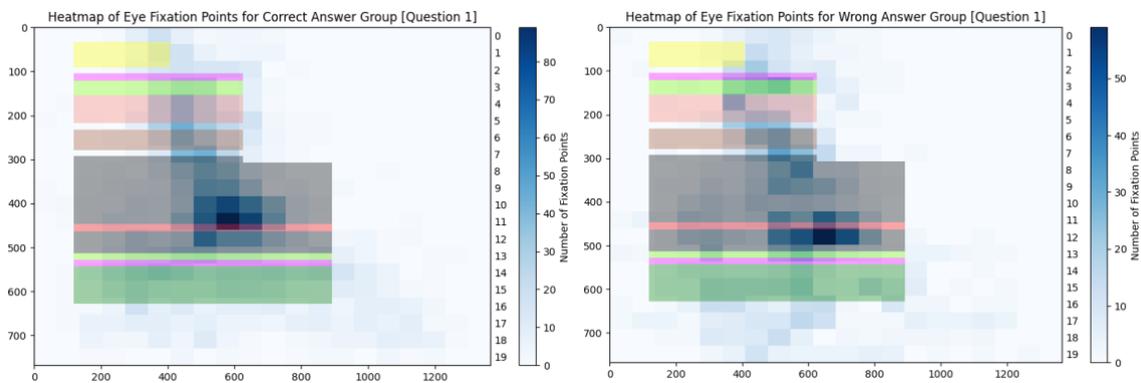

*Figure 1.* The eye fixation points heatmap graph of right and wrong answer groups for Exercise 1.

### 3.2 Visual Cumulative Fixation Point Graph of Exercise 3

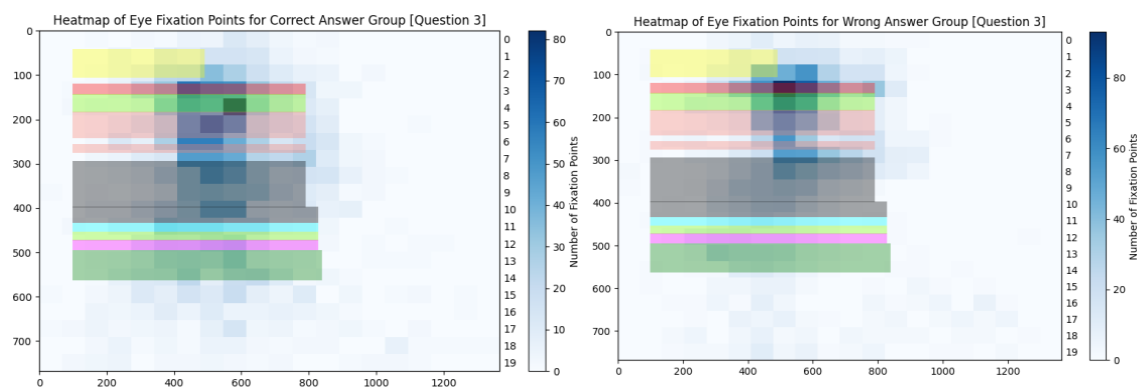

*Figure 2.* The eye fixation points heatmap graph of right and wrong answer groups for Exercise 3.

### 3.3 Visual Cumulative Fixation Point Graph of Exercise 5

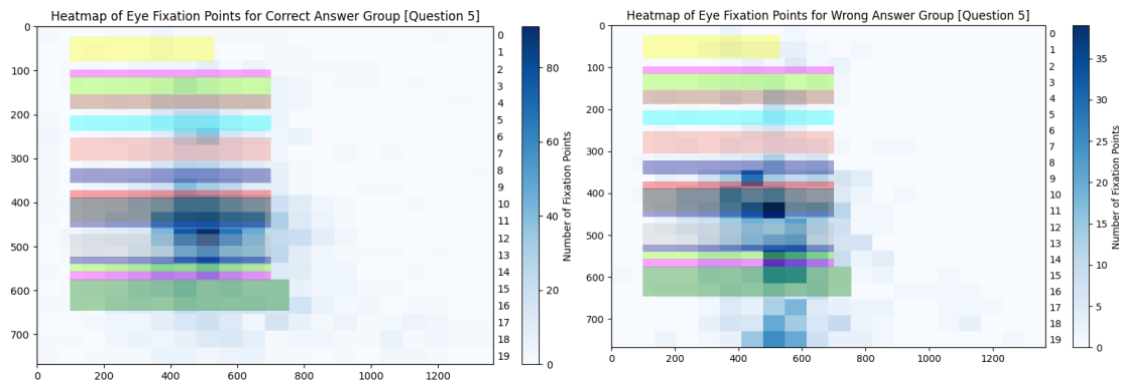

*Figure 3.* The eye fixation points heatmap graph of right and wrong answer groups for Exercise 5.

Table 1. *Mann-Whitney U Test on the Correct and Wrong Answer Groups Per Exercise.*

|                | Exercise 1 | Exercise 3 | Exercise 5 |
|---|---|---|---|
| Test statistic | 80,468.0   | 91,196.5   | 90,522.5   |
| p-value        | 0.8822     | **0.0004**\*\* | **0.0007**\*\* |

\*\*p is less than 0.001

### 4. Conclusion and Acknowledgement

Each exercise featuring the two groups have almost similar cumulative eye fixation distribution via their heatmaps. It provides evidence that the student may look at AOIs near the error line of the code but may flag the incorrect line while debugging. According to the p-value of the Mann-Whitney U statistic, the visual graph of Exercise 1 supports that there is no deviation on the performance between the two cohorts. While Exercise 3 and 5 suggest significant difference on the cognitive strategy of how the two groups answered the programming exercise. Both arguments are indicators that the eye-mind link hypothesis may not necessarily be exhibited whenever debugging program code. Future research should consider a numerical quantity on how similar the heatmaps are. Additionally, more participants from other universities must be included to dilute biases of the sample enrolled in the same institution.